\documentstyle[new_cite,times,epsf,psfig]{mn}

\title{X-ray spectral variability of the Seyfert galaxy NGC 4051}
\author[]
  {G.~Lamer $^1,^2$,\thanks{E-mail: glamer@aip.de} 
  I.M.~M$^{\rm c}$Hardy $^1$,  P.Uttley $^1$ and K. Jahoda$^3$\\
  $^1$Department of Physics and Astronomy, The University,
       Southampton, SO17 1BJ \\
  $^2$ Astrophysikalisches Institut Potsdam, Germany \\
  $^3$Laboratory for High Energy Astrophysics, Goddard Space Flight Center} 

\date{Accepted. Received}
\pagerange{\pageref{firstpage}--\pageref{lastpage}}
\pubyear{2002}

\pagerange{\pageref{firstpage}--\pageref{lastpage}}
\pubyear{2002}

\begin{document}

\maketitle

\label{firstpage}

\begin{abstract}

We report on the X-ray spectral variability of the Seyfert 1 galaxy
NGC~4051 observed with the Rossi X-ray Timing Explorer ({\em RXTE})
during a 1000 day period between May 1996 and March 1999.  The spectra
were obtained as part of monitoring observations and from two long
observations using the {\em RXTE} Proportional Counter Array (PCA).
During the monitoring period the 2-10 keV flux of NGC~4051 varied
between $10^{-12}$ and $7\cdot 10^{-11} {\rm erg/(cm^2\;s)}$.  We
re-analysed {\em RXTE} PCA observations from a distinct low state in
May 1998 using the latest background and detector response models.
The {\em RXTE} and {\em BeppoSAX} observations of NGC~4051 during the
low state show a very hard spectrum with a strong unresolved fluorescence
line.  This emission, probably due to reflection from a molecular
torus, is likely to be constant over long time-scales and is therefore
assumed as an underlying component at all flux states.  By subtracting
the torus component we are able to  determine the spectral
variability of the primary continuum.  In the variable component we
observe a strong anti-correlation of X-ray flux and spectral hardness
in the PCA energy band. We show that the changes in hardness are
caused by slope variability of the primary power law spectrum rather
than by changing reflection or variable photoelectric absorption. The
primary spectral index varies between $\Gamma=1.6$ for the faintest
states and $\Gamma=2.3$ during the brightest states, at which level
the spectral index approaches an asympotic value.
%This is a broader range than
%predicted by pair-dominated Comptonisation models.
We find that the
response of the flux of the 6.4 keV iron fluorescence line to changes
in the continuum flux depends on the timescale of the observation.
The profile of the line is very broad and indicates an origin in the
innermost regions of the accretion disk.

\end{abstract}


\section{Introduction}

It is thought that the central engines of active galactic nuclei (AGN) 
are powered by the infall of matter onto a supermassive black hole. 
Theoretical arguments suggest the formation of an accretion disk
which extends from the innermost stable orbit to an outer radius on the 
scale of light days.
The popular unified scheme for Seyfert galaxies (eg Antonucci 1993)
requires a dusty torus, probably coplanar with the accretion
disk, which is opaque to radiation from the near infrared to soft X-rays.  
According to the unified scheme Seyfert type 1 and  Seyfert type 2 
AGN are intrinsically identical object classes with the only
difference being that in Seyfert type 2 galaxies, due to their different 
orientation relative to the observer, the  central source 
and the broad line region (BLR) are obscured by a dusty torus. 
The first probable direct detection of a torus was shown in our earlier
SAX \cite{Guainazzi} and RXTE \cite{Uttley99} observations
of NGC~4051.  
However, in general, detection of the torus and
the determination of its geometry has not been possible.   

X-ray observations have shed some light on the structure of the
innermost regions of AGN. The soft X-ray spectrum 
often has a relatively steep slope, which has been explained by 
thermal emission from an accretion disk. However, there are 
alternative theoretical interpretations and 
due to the limited spectral resolution in this energy range
the soft X-ray spectra are poorly understood.   
    
At higher energies, a hard power law spectrum dominates  the emission. 
This component is probably due to Comptonization of the thermal
UV photons in a hot  corona surrounding the disk.

The averaged {\it Ginga} medium energy X-ray spectrum of several
Seyfert galaxies showed the presence of an emission line at 6.4 keV,
an absorption feature at 7-8 keV, and a further flattening of the
spectrum beyond $\sim 10$ keV.  These features are interpreted as iron
$K_{\alpha}$ emission, iron K edge absorption and Compton reflection
and therefore are evidence for reprocessing of X-rays by relatively
cold matter, The discovery of broadening due to gravitational redshift
and Doppler shifts of the iron K$_{\alpha}$ fluorescence lines in
Seyfert X-ray spectra suggests that at least part of the reprocessing
takes place in the inner parts of the accretion disk ( \ncite{Tanaka},
\ncite{Nandra97}, \ncite{Reynolds}).  However, the detection of narrow
components to the Fe K$_\alpha$ line with {\em ASCA}
(eg \ncite{Weaver97}) and recently {\em Chandra} (\ncite{Yaqoob}, \ncite{Kaspi})
and {\em XMM-Newton} \cite{Reeves}
suggests reprocessing of X-rays in matter at a larger distance from
the black hole, eg in the BLR or in the molecular torus.

The analysis of X-ray variability is a powerful tool for the
investigation of the inner regions of the AGN central engine, since
the time-scales of the variability can give an indication of the
geometrical sizes of the regions involved.  \scite{Yaqoob96} reported
a rapid ($<3\cdot 10^4$ s) response of the Fe fluorescence line flux
to the continuum flux in NGC~7314. From the resulting maximum distance
of the reflecting material from the primary X-ray source they derive
an upper limit on the mass of the black hole.  
% I presume they must use some velocity? eg line width?...Ian
However, tdhe analysis of fluorescence line variability in a number of
other Seyfert galaxies have led to confusing and partly contradictory
results. While in some sources a positive correlation of line and
continuum flux has been observed, averaged over long timescales,
(eg NGC~5506, \ncite{Lamer2000}), other studies found no evidence
for a close relationship between continuum and fluorescence line
(\ncite{Weaver2001}, \ncite{Vaughan}).

Flux-dependent variability of the continuum spectral shape itself can be 
used to constrain models of the Comptonising accretion disk corona (eg
\ncite{Haardt}).  In general, the continuum slope
appears to be correlated with flux (eg \ncite{Leighly}; \ncite{mch98} ;
\ncite{Lamer2000}), as predicted by
most simple Comptonisation models.  However, it is important to
determine whether the degree of slope variability, and the form of the
correlation with flux (eg does the continuum slope saturate at some 
maximum value?) are in agreement with existing models.

NGC~4051 is a nearby (z=0.0023) low luminosity Seyfert 1 galaxy, which
is among the most variable AGN in the X-ray band \cite{Green}.  We
have been monitoring NGC 4051 with RXTE since 1996.  Since our
monitoring commenced, the 2-10 keV flux from the object has varied by
a factor of $\sim$100. In 1998 NGC~4051 entered a state of extremely
low X-ray flux that lasted for $\sim 150$ days (\ncite{Guainazzi},
\ncite{Uttley99}).  Here we report on the spectral changes that
accompanied the dramatic flux variability. In section \ref{obs} we
describe the observations and the reduction of the {\em RXTE} data.
Results from observations in May 1998, during the extreme low state,
are re-analysed in a consistent manner with the other observations and are
discussed in section \ref{lowstate}.  In section \ref{specvar} we
present the analysis of the continuum and fluorescence line variations
observed during the long term monitoring and during an {\em RXTE} long
look in December 1996.

\section{Observations and data reduction}
\label{obs}

We have been monitoring NGC 4051 with RXTE since May 1996.  The
observations were separated by a range of time intervals.  For the
first 6 months the observations took place weekly with periods of
twice-daily observations in May 1996 and daily observations in October
1996. Since November 1996 the source has been observed at fortnightly
intervals.  {\it RXTE} observed NGC 4051 with the Proportional Counter
Array (PCA) and the High Energy X-ray Timing Experiment (HEXTE). The
PCA \cite{Zhang} consists of 5 Xenon-filled Proportional Counter Units
(PCUs), sensitive to X-ray energies from 2-60 keV. The maximum
effective area of the PCA is 6500 cm$^2$.  As the signal to noise
ratios of the higher energy HEXTE spectra on NGC 4051 are low, we only
present the results from the PCA.  For technical reasons the high
voltage settings of the PCA units have been adjusted several times
during their lifetime. In this paper we only present data taken during
the PCA gain epoch 3, which lasted from April 1996 to March 1999. The
total exposure time of the 130 individual monitoring observations
during gain epoch 3 is 85 ksec.  Apart from the monitoring
observations {\it RXTE} performed two long-look observations of NGC
4051 in December 1996 with a total on-source time of 36.4 ksec within
3 days and in May 1998 for 84.4 ksec (on-source) within 2 days.

We have used  {\sc ftools v4.2} for the reduction of the PCA and HEXTE
data. PCA ``good times'' have been selected from the Standard 2 mode data
sets using the following criteria: target elevation $> 10^{\circ}$,
pointing offset $<0.01^{\circ}$, time since SAA passage $> 30$ min,
standard threshold for electron contamination.
Occasionally one or more of the 5 Proportional Counter Units (PCUs) of
the PCA were switched off during the observations. 
In order to maximise the signal to noise ratio of the spectra, we 
used the data from all PCUs that were switched on during each
individual pointing and used only PCU layer 1. 
We calculated the background in the PCA with the tool 
{\sc pcabackest v2.1} using the L7 model for faint sources, which 
is  suitable for determining the PCA background for energies $\leq 24$ keV.
From a series of $\sim 100$ blank field observations from the public archive
(proposal ID  P30801) we determined the uncertainties in the estimated
background rates as a function of energy. At energies below 24 keV the 
1$\sigma$ uncertainty of the background is of the order  
$0.01 {\rm cts\; s^{-1} keV^{-1}}$ (see \ncite{Lamer2000} for details).
The PCA response matrices were calculated invididually for each
observation using {\sc pcarsp v2.37}, taking into account temporal
variation of the detector gain and the changing numbers of detectors
used.

 \begin{figure*}
 \par\centerline{\psfig{figure=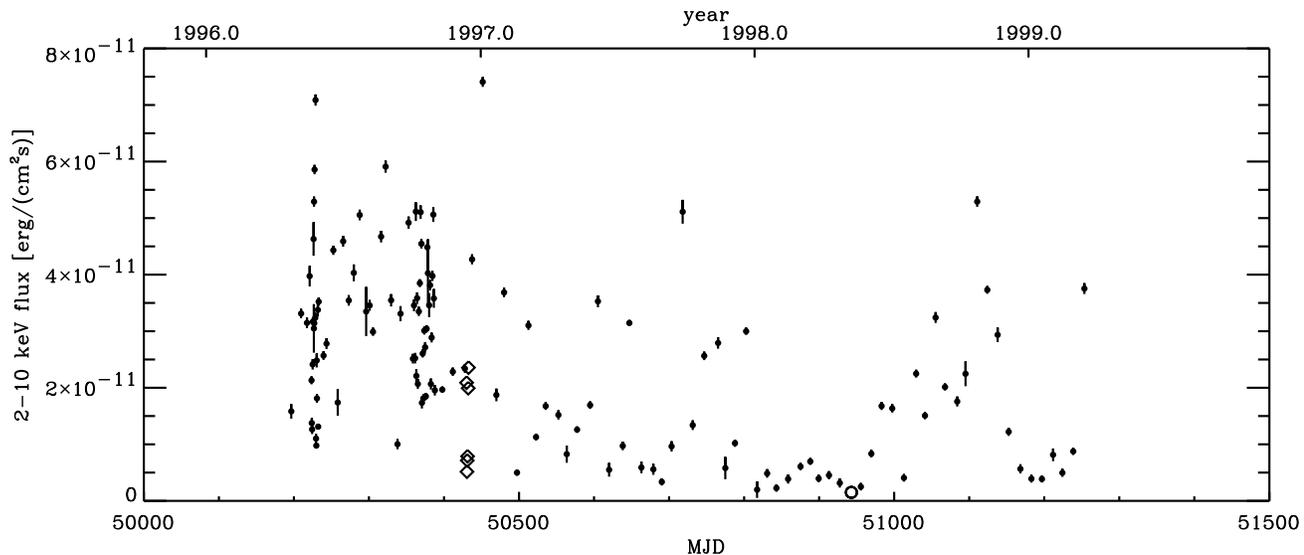,width=17truecm}}
 \caption{\label{lightcurve}
 2-10 keV lightcurve of NGC 4051 during the long term 
 monitoring campaign (filled circles with error bars). The diamonds
 show the lightcurve from  the December 1996 observation, the open
 circle indicates the flux level during the May 1998 observation at the
 end of the 150 day long low state.}
 \end{figure*}

\section{The low state spectrum}
\label{lowstate}

\begin{table*}
\centering
\caption{\label{specfit} Fit parameters for the May 1998 low state spectrum}
\begin{tabular}{ @{}lcccccc@{} }
model     & $\Gamma$& f$_{6.4}$   & EW$_{6.4}$  &  EW$_{7.057}$& EW$_{7.477}$  & $\chi^2$ (dof) \\
          &         & $10^{-5} \rm{ph\;s^{-1} cm^{-2} }$  & [keV]& [keV]& [keV] &               \\
[10pt] 
\multicolumn{6}{c}{RXTE PCA}\\
[10pt] 
{\sc pwl+zgauss}    & $0.56\pm0.11$  &$2.80\pm0.70$ & $1.40\pm0.35$ & -    & -    & 38.8 (42) \\
{\sc pwl+3 zgauss}  & $0.52\pm0.11$  &$2.57\pm0.53$ & $1.40\pm0.34$ & 0.16 & 0.10 & 36.8 (42) \\
{\sc pexrav+zgauss} & $2.02\pm0.10$  &$2.16\pm0.75$ & $0.79\pm0.35$ &  -   & -    & 39.8 (42) \\
{\sc pexrav+3 zgauss}& $2.01\pm0.10$ &$2.35\pm0.66$ & $0.88\pm0.27$ & 0.12 & 0.11 & 38.0 (42) \\
[10pt] 
\multicolumn{6}{c}{BeppoSAX MECS}\\
[10pt] 
{\sc pwl+zgauss}   & $0.37\pm0.17$  & $ 1.88\pm0.60$& $1.16\pm0.37$ & -    &   -  & 28.4 (29) \\
{\sc pwl+ 3 zgauss}& $0.52\pm0.18$ & $ 2.04\pm0.64$& $1.26\pm0.39$ & 0.14 & 0.09 & 25.5 (29) \\  
{\sc pexrav+zgauss}& $1.75\pm0.18$ &  $ 1.85\pm0.64$& $0.87\pm0.30$ & -    &   -  & 33.9 (29) \\
{\sc pexrav+3 zgauss}& $1.83\pm0.20$ & $ 1.12\pm0.63$& $0.50\pm0.28$ & 0.07 & 0.06 & 29.6 (29) \\

\end{tabular}
\end{table*}

Our monitoring with the RXTE PCA revealed that during a 150 day period
from January 1998 to May 1998 NGC~4051 was in a state of unusually low
X-ray flux (\ncite{Uttley99}, see Fig. \ref{lightcurve}).  Near the end
of this period NGC~4051 was observed simultaneously by {\it RXTE}
\cite{Uttley99} and {\it BeppoSAX } \scite{Guainazzi}.  We derived a
2-10 keV flux of $2.0 \cdot 10^{-12} \rm{erg s^{-1} cm^{-2} }$ from
the {\it RXTE} PCA spectrum and $1.4 \cdot 10^{-12} \rm{erg s^{-1}
cm^{-2} }$ from the {\it BeppoSAX } MECS data.  Both observations
showed a very hard X-ray spectrum and unresolved 6.4 keV iron line
emission with a very high equivalent width ($\sim 1000$ eV).  This
spectrum was interpreted as a pure reflection component with the
central source  being virtually switched off. Since the reflected
emission was still present after the source had been in a low state
for 150 days, the distance of the reflector from the central source
must be $\geq$ $10^{17}$ cm.  It is likely that the Compton reflection
and iron line fluorescence from the distant reflector is a constant
contribution to the X-ray spectrum of NGC 4051.  For the investigation
of the variability of the primary X-ray source in NGC 4051 it is
necessary to take this underlying component into account.

We have therefore re-analysed the PCA data from the low state using
the latest response matrices, appropriate to gain epoch 3, and
background models.  We fitted two models to the PCA and {\it BeppoSAX}
MECS spectra: a simple power law model with a Gaussian emission
line and a pure reflection model, also with a Gaussian emission line.
The model for a pure reflection component was calculated by using the
{\sc XSPEC PEXRAV} model (\ncite{MZ}) with a fixed reflected fraction
$R=100$.  With this setting the contribution of the intrinsic power
law emission is negligible. See Table~\ref{specfit} for the results
of the spectral fits. Throughout the paper 1~$\sigma$ confidence
limits are given as error values. The error of the fit parameters in
Table~\ref{specfit} were calculated for 4 interesting parameters.
Due to the relatively low signal-to-noise ratios of the data a clear
cut decision between these two models is not possible.  For both the
PCA and MECS spectra the power law model yields a slightly better
$\chi^2$ value than the pure reflection fit.  However, the power law
fit requires a photon index of $\Gamma\sim0.5$, a value that is
unlikely for standard Comptonization models but which might be
applicable to an advective flow. If the low state X-ray
emission solely arises from reflection, the {\sc PEXRAV} model fit to
the PCA spectrum results in a photon index $\Gamma\sim2.0$ for the
primary emission. Assuming that the reflecting matter subtends a solid
angle of 2$\pi$ steradians or less, the required incident radiation
corresponds to a 2-10 keV flux level of $\geq 2\cdot10^{-11} {\rm erg
s^{-1} cm^{-2}}$. This is in good agreement with the flux range of
$\sim 1\cdot10^{-11}$ to $\sim 6\cdot10^{-11} {\rm erg s^{-1}
cm^{-2}}$ found during the monitoring of NGC~4051 before the 150 day
low state period.
 
The strong iron $K_{\alpha}$ fluorescence line supports the
interpretation that the low state spectrum is a pure reflection
component.  The very low level of the continuum in these observations
of NGC~4051 allow a more detailed analysis of any fluorescence or
emission lines.  When fitting a power law or {\sc PEXRAV} model with a
single Gaussian line to the PCA spectrum, the resulting source frame
line energy of $\sim$ $6.6\pm0.1$ keV is slightly higher than the
nominal value of 6.4 keV for fluorescence from cold iron. The best fit
model to the {\it BeppoSAX} MECS spectrum includes a narrow line at
$6.47\pm0.06$ keV, consistent with 6.4 keV. The fit residuals,
however, show some excess flux at $\sim$ 7 keV.  When fixing the
energy of the fluorescence line to 6.4 keV, excess flux at $\sim$ 7
keV is evident in all spectra.  We therefore included narrow (0.05
keV) iron K$_\beta$ and nickel K$_\alpha$ fluorescence lines at 7.058
keV and 7.477 keV into the spectral model.  The fluxes of these
additional lines were set to Fe K$_\beta$/Fe K$_\alpha$ = 0.11 as
theoretically expected for neutral iron \cite{Kikoin} and Ni
K$_\alpha$/Fe K$_\alpha$ = 0.07 as expected for solar abundances
\cite{Nandra99}.  Taking these lines into account slightly improves
the $\chi^2$ of both the PCA and the MECS fits without increasing the
number of free parameters (see Table~\ref{specfit}).
 
\begin{figure}
 \par\centerline{\psfig{figure=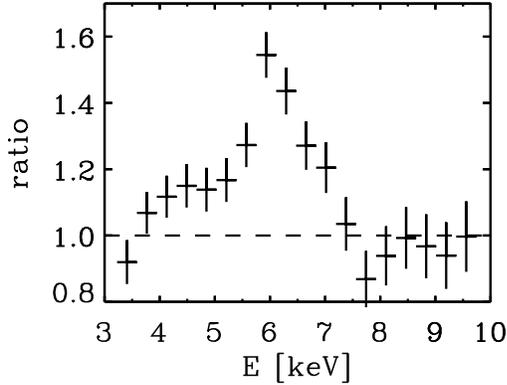,width=8truecm}}
 \caption{\label{line_dec}
 Fluorescence line profile in the lowest flux state during the December 
 1996 observation. The flux of the narrow component is consistent with
 the line flux during the May 1998 low state.}
 \end{figure}

Throughout the remainder of this paper, we will assume that a narrow
iron line component, whose flux is consistent with the line flux 
measured in May 1998, is present in
the spectrum of NGC~4051 over the 3 years of our RXTE observations. 
Fig. \ref{line_dec} shows the Fe ${\rm K}{\alpha}$ line profile with a possible
strong narrow component during  the lowest flux  states of the
December 1996  observation (see first row of Table~\ref{fluxbin_dec}).
This is consistent with our view that a component, 
which is constant on the time-scale of years  is present in the X-ray 
spectrum of NGC~4051.   
In the analysis of the spectral variability in the {\em RXTE} PCA
data, we therefore add the {\sc PEXRAV + 3 ZGAUSS} model with the 
parameters  derived from the PCA low state data (see Table~\ref{specfit}) 
as a constant component to all model fits of the 
monitoring and December 1996 data.

We note that the presence of narrow line components in some other
objects has been confirmed by {\it Chandra} and {\it XMM}
observations. \scite{Yaqoob} detected a narrow line at 6.40 keV in a
{\it Chandra} High-Energy Transmission Grating spectrum of
NGC~5548. Since the line is slightly resolved, they assume an origin
in the broad line region. 
\scite{Ogle} suggested that part of the narrow iron line 
observed by {\it Chandra} in NGC~4151 may originate in the narrow-line region. 
\scite{Reeves} found an unresolved component at 6.4 keV in the {\it XMM} EPIC 
spectrum of the low-luminosity quasar MKN~205. \scite{Kaspi} were able
to resolve  the narrow 6.4 keV line
in NGC~3783 with the {\em Chandra} HETGS. 
The velocity dispersion of FWHM=1720 ${\rm km s^{-1}}$
suggests that the line is emitted in the outer BLR or the inner part of the molecular torus.

Although we can say little about the line width from our RXTE
observations, the fact that the line remains detectable in the very
low state May 1998 observations, 150 days after the continuum went
into the low state, means that it could not come from the broad line
region, but part of it might come from the narrow line region. However
the very hard continuum is not easily explained by scattering from the
narrow line region.

If the strong torus contribution to the iron fluorescence line in
NGC~4051 is common in other Seyfert galaxies, some of the previous
results on disk line spectroscopy will have to be revised, once high
resolution spectra become available.

\section{Spectral variability}
\label{specvar}

The large variations in the X-ray flux observed within the 3 years of
monitoring of NGC~4051 make the RXTE observations ideally suited for
an investigation of associated spectral variability.  Since the signal
to noise ratio in any one of the individual 1 ksec pointings is not
sufficient for individual spectral analysis, we have grouped and added
the spectra according to source flux.  Since the data were gathered by
different combinations of PCUs and the detector gains changed slightly
over time, we used fitted model fluxes for the grouping of the data.
In order to obtain the flux level for each spectrum we fitted a
single power law model to each spectrum in the 2-10 keV range using
the appropriate detector response for each spectrum. The monitoring
observations were grouped into 7 flux bins, while the spectra from the
December 1996 long look observation were grouped into 4 bins (see
Tables~\ref{fluxbin_mon} and \ref{fluxbin_dec} for the boundaries of
the bins). The appropriate detector response matrices for each
spectrum have been created by averaging the response matrices of the
individual input spectra (using their integration times as weighting
factors).

As described in section \ref{lowstate} we regard the emission observed
during the low state in May 1998 as a constant underlying spectral
component from a distant reflector.  For all spectral fits in this
section we have therefore included the best fit of the {\sc PEXRAV + 3
ZGAUSS} model to the May 1998 PCA data as a fixed additional
component.

\begin{table}
\centering
\caption{\label{fluxbin_mon}Flux bins for monitoring observations }
\begin{tabular}{rcr}
flux range & mean flux & exposure \\
\multicolumn{2}{c}{[$\rm{erg\; s^{-1}\; cm^{-2}}$] } & [sec]    \\[10pt]
$<1.0\cdot10^{-11}$    & $5.47\cdot10^{-12}$      & 21824  \\
$1.0-2.0\cdot10^{-11}$ & $1.63\cdot10^{-11}$      & 22352  \\
$2.0-3.0\cdot10^{-11}$ & $2.76\cdot10^{-11}$      & 15744  \\
$3.0-3.5\cdot10^{-11}$ & $3.36\cdot10^{-11}$      &  8288  \\
$3.5-4.5\cdot10^{-11}$ & $4.06\cdot10^{-11}$      & 10880  \\
$4.5-5.5\cdot10^{-11}$ & $5.21\cdot10^{-11}$      &  4176  \\
$>5.5\cdot10^{-11}$    & $6.66\cdot10^{-11}$      &  2016  \\
\end{tabular}
\end{table}

\begin{table}
\centering
\caption{\label{fluxbin_dec}Flux bins for December 1996 observation }
\begin{tabular}{rcr}
flux range & mean flux & exposure \\
\multicolumn{2}{c}{[$\rm{erg\; s^{-1}\; cm^{-2}}$] } & [sec]    \\[10pt]
$<1.0\cdot10^{-11}$     & $6.78\cdot10^{-12}$      & 20800  \\ 
$1.0-2.0\cdot10^{-11}$ & $1.53\cdot10^{-11}$      & 7792   \\
$2.0-3.0\cdot10^{-11}$ & $2.65\cdot10^{-11}$      & 6800   \\
$>3.0\cdot10^{-11}$     & $3.62\cdot10^{-11}$      & 1008   \\
\end{tabular}
\end{table}

\subsection{Continuum and reflection}

Simple power law fits to the 
flux binned spectra from the monitoring
observations after removal of the torus component still 
reveal a strong correlation of the spectral slope 
with the flux level of NGC~4051: the softest spectra are found 
when the flux is highest (see fig \ref{lineflux}).
In the context of the standard  models for the X-ray emission of
Seyfert galaxies three principal mechanisms can contribute to this
correlation: 

\begin{itemize}
  \item[1.] Intrinsic spectral variability of the primary continuum. 
            During states of high luminosity the higher flux of 
            UV/soft X-ray photons may cool the accretion disk corona
            and therefore lead to a softer Compton spectrum at higher 
            X-ray energies.
  \item[2.] The fraction of reflected radiation may be higher during
            periods of low luminosity (eg due to an additional, 
            non-variable reflection component). 
  \item[3.] Gas near the X-ray source that is usually 
            strongly ionised by radiation from the central
            source might recombine to less ionised states when the
            luminosity of the AGN drops.  The resulting increase in
            absorption of soft X-ray then leads to a hardening of the spectrum.
            Outside of the very low luminosity states
            NGC~4051 is observed to have a significant ``warm absorber''
            with a column density of 
            $5-8 \cdot10^{22} {\rm cm^{-2}}$ (\ncite{imh95}, \ncite{Komossa})
            which could lead to substantial absorption of soft X-rays if
            the warm absorbing gas recombined.
\end{itemize}

\begin{figure}
 \par\centerline{\psfig{figure=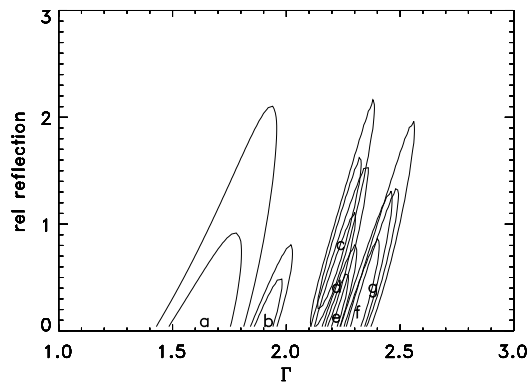,width=8truecm}}
 \caption{\label{mon_cont}
 68\% and 90\% confidence contours of the parameters photon index
 $\Gamma$ and  reflected fraction $R$ for the {\sc PEXRAV} model fits
 to  the flux selected PCA spectra. The letters 'a' to 'g' denote
 the spectra in the order of increasing flux levels. See section           
 \ref{specvar} for a full description of the model fits.}
 \end{figure}

In order to distinguish between the first two possibilities we fitted
{\sc PEXRAV} models with a variable Gaussian line in addition to the
fixed torus reflection model to the flux selected spectra of both the
monitoring and December 1996 observations.  We reduced the number of
free parameters by fixing some of the less critical parameters of the
{\sc PEXRAV} model to reasonable values. The high energy cutoff of the
primary power law was set to 300 keV, effectively eliminating the
cutoff.  The iron and metal abundances were set to the solar values
and the inclination angle was fixed at $30^\circ$ as suggested by the
results of disk line model fits to ASCA spectra \cite{Nandra97}.
Confidence contours of the primary photon index $\Gamma$ and the
reflected fraction $R$ are shown in Figs~\ref{mon_cont} and
\ref{dec_cont}. It is evident that a continuum of variable spectral
index is needed to satisfy the data. The best fit photon indices range
from $\Gamma=1.6$ for the lowest flux states to $\Gamma=2.3$ for the
highest flux states of the monitoring observations. 
Note that the inclusion of the presumed constant torus component
does not change the results qualitatively. If we omit this fixed 
component, the photon index of the lowest flux spectrum is  $\Gamma=1.2$;
the fit to the highest flux spectrum is not affected by this change. 
No variability of
the reflected fraction is obvious and at the $90\%$ confidence levels
all spectra are consisistent with zero reflection.  Apart from the
fact that the constant reflected (torus) component will become less
prominent at higher flux levels, there is no obvious contribution of
reflection to the spectral variability.

We also investigated the warm absorber as the third possible source of
hardness variability.  Again including the low state spectrum as a
constant component, we modelled the flux-selected X-ray spectra with a
power law and Gaussian emission line absorbed by a warm absorber ({\sc
XSPEC ABSORI} model) with a column density of $N_{\rm H} =
5\cdot10^{22} {\rm cm^{-2}}$, as suggested by \scite{Komossa}.  The
models were calculated on a grid in spectral index $\Gamma$ and
ionisation parameter $\xi$ and fitted to each of the 7 flux selected
spectra of the monitoring observations.  As a result, the ionisation
parameter appears to be poorly constrained.  The spectral index is
clearly correlated with flux and there is no $\Gamma$ that would be
consistent with all of the 7 spectra (see Table \ref{warmabs}).  We
therefore conclude that changes in the ionisation state of the warm
absorber in NGC 4051 do not play a significant role in the spectral
variability of the source.

\begin{table}
\centering
\caption{\label{warmabs}Results of warm absorber fits to flux 
selected monitoring spectra}
\begin{tabular}{lcl}
  mean flux               & photon index &  $\xi$ \\
$[\rm{erg\; s^{-1}\; cm^{-2}}]$ & 1$\sigma$ range &  1$\sigma$ limit \\[10pt]
 $5.47\cdot10^{-12}$      &  1.50-1.91   & $> 0$  \\
 $1.63\cdot10^{-11}$      &  1.85-2.14   & $> 0$  \\
 $2.76\cdot10^{-11}$      &  2.10-2.30   & $> 60$ \\
 $3.36\cdot10^{-11}$      &  2.13-2.28   & $> 120$\\
 $4.06\cdot10^{-11}$      &  2.08-2.28   & $> 370$\\
 $5.21\cdot10^{-11}$      &  2.28-2.43   & $> 170$\\
 $6.66\cdot10^{-11}$      &  2.29-2.45   & $> 240$\\
\end{tabular}
\end{table}
  
To summarize, variability of primary power law continuum slope is the
only viable explanation of the strong hardness variability in the 2-24
keV X-ray spectrum of NGC 4051.

\subsection{Fluorescence line variability}

 \begin{figure}
 \par\centerline{\psfig{figure=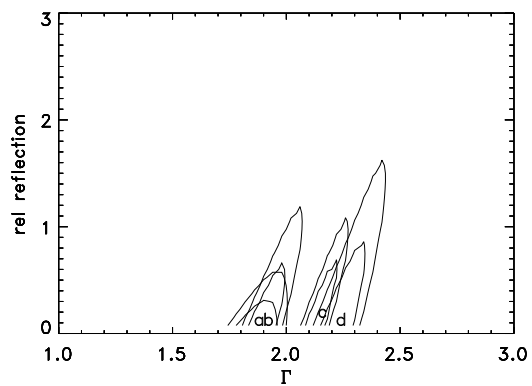,width=8truecm}}
 \caption{\label{dec_cont}
 Same as Fig. \ref{mon_cont} for the flux selected spectra of the
 December 1996 observation. The letters 'a' to 'd' denote the 4 flux
 levels in ascending order. 
 }
 \end{figure}

 \begin{figure}
 \par\centerline{\psfig{figure=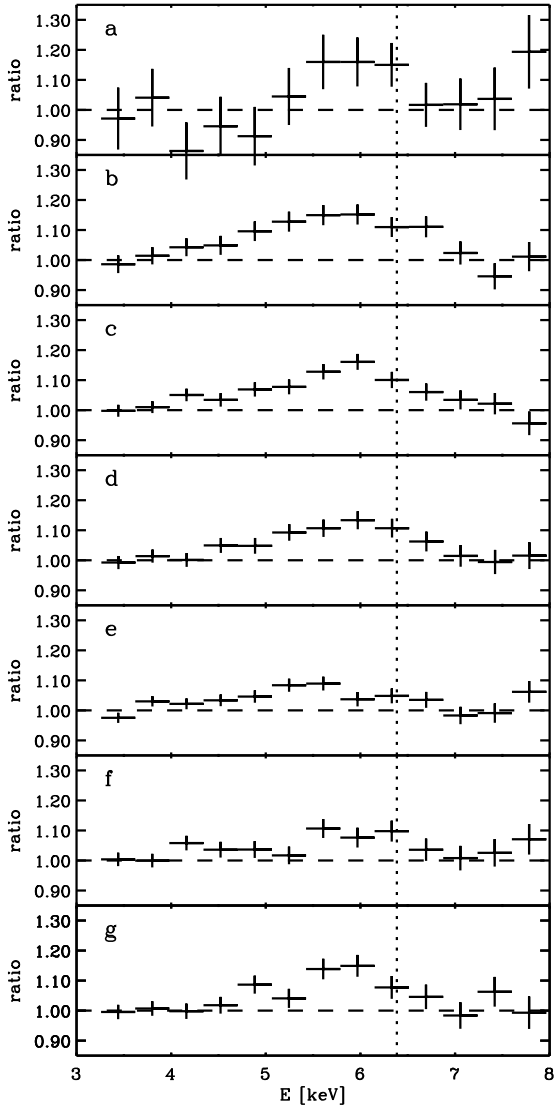,width=8truecm}}
 \caption{\label{monline}
 Iron line profiles in the flux selected spectra of the monitoring
 observations. The letters a-g denote the 7 flux intervals 
 in ascending order. The low state spectrum including the narrow
 6.4 keV fluorescence line has been subtracted. 
 Note the strong broadening of the line at all flux levels.
 }
 \end{figure}

 \begin{figure}
 \par\centerline{\psfig{figure=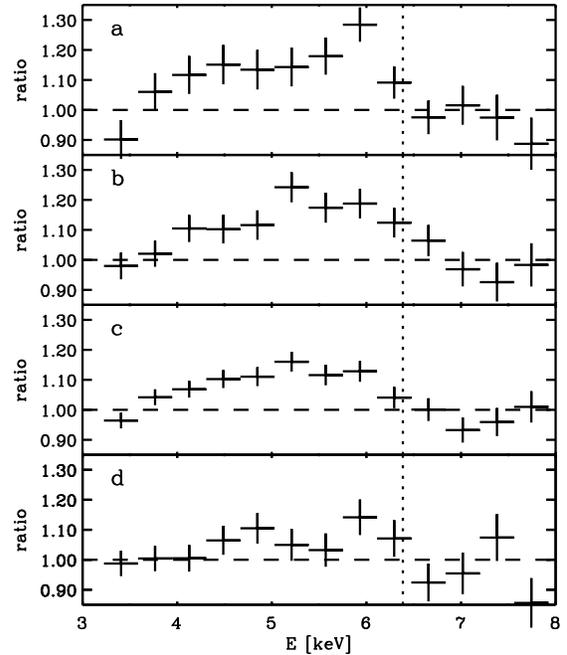,width=8truecm}}
 \caption{\label{decline}
 Iron line profiles for the December
 1996 observations in the 4 different flux ranges. 
 As in Figure \ref{monline} the low state spectrum
 including the narrow line has been subtracted.
 }
 \end{figure}

After subtraction of the low state reflection spectrum with the narrow
iron fluorescence line, broad fluorescence line emission is still
detectable at all flux levels.  Figs. \ref{monline} and \ref{decline}
show ratio plots obtained by fitting power law models to the flux
selected spectra excluding the 4-8 keV spectral range.  The line is
broad and strongly redshifted as expected for an origin in the inner
parts of an accretion disk.

We fitted each of the flux selected spectra with a model consisting of
the best fit low state spectrum as discussed in section
\ref{lowstate}, a power law, and the {\sc XSPEC DISKLINE} model. The
{\sc DISKLINE} model calculates the profile of a line originating from
a Keplerian disk surrounding a Schwarzschild black hole.  The rest
energy of the line was fixed at 6.4 keV.  The geometric origin of the
fluorescence line is defined by three parameters: the inner radius
$R_{\rm i}$, the outer radius $R_{\rm o}$ and the power law index $q$
which describes the radial variation of the emissivity.  We set
$R_{\rm i}$ to 6 $r_{\rm g}$, the radius of the innermost stable orbit
for a Schwarzschild black hole.  The outer radius $R_{\rm o}$ was set
to 400 $r_{\rm g}$.  The emissivity index $q$ was left free to vary.

Table \ref{diskline} summarizes the results of the disk line
modelling. For each spectrum we give both the equivalent widths of the
narrow fluorescence lines from the fixed narrow line component and the
best fit values for the broad disk line.  The values for the narrow
lines are derived from the total fluxes of the lines at 6.4 keV, 7.06
keV, and 7.477 keV (see section \ref{lowstate}).  Note that the narrow
line fluorescence from the torus can be a major contribution to the
total line flux, in particular during lower flux states. Hence the
profile of the disk line can only be determined if the contribution
from the torus can be reliably measured.  The best fit values of the
disk inclination confirm the values of $i\sim 30^{\circ}$ resulting
from {\em ASCA} observations \cite{Nandra97}.  The values of $q$ in
most cases are poorly constrained, but generally the emissivity drops
steeper than with $q=-3$.  This indicates a concentration of the line
emission in the innermost regions of the disk, as suggested by the
broad profile of the iron line.  Since the {\sc PEXRAV} model fits
described in section \ref{specvar} required no disk reflection (see
Figs \ref{mon_cont}+\ref{dec_cont}) we do not include disk reflection
in the disk line model fits.  However, we repeated the model fits
using a {\sc PEXRAV} model with reflected fraction $R=1$ instead of
the power law model.  Including reflection to the fits did not change
the line fluxes or disk line parameters significantly.

\begin{figure}
 \par\centerline{\psfig{figure=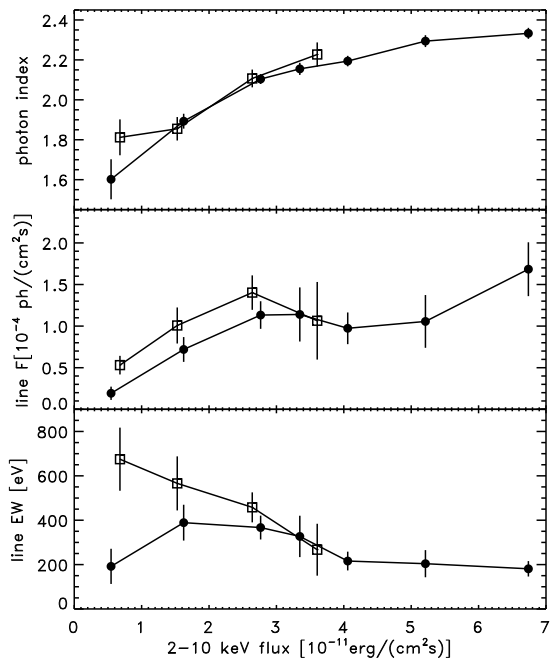,width=8truecm}}
 \caption{\label{lineflux}
 Flux dependencies of  power law spectral index (top), disk line
 flux (middle) and disk line equivalent width (bottom). The results 
 from the monitoring observations are plotted as dots, the December 
 1996 long look is represented by  open squares. Error bars indicate
 $\Delta\chi^2=1.0$. 
 }
 \end{figure}

The best fit photon indices, line fluxes, and line equivalent widths
are plotted in Fig. \ref{lineflux}.  The best fit values of the
fluorescence line flux from the long term monitoring observations show
a clear correlation with the 2-10 keV flux. 
However, the correlation is not
linear, in the continuum flux range range $(2.0..5.5) 10 ^{-11} {\rm
erg\; cm^{-2} s^{-1}}$ the line flux is nearly constant.  The variation
of the equivalent width is not very pronounced, the lowest and the
three highest flux bins are consistent with $EW=200{\rm eV}$.

On the shorter time scales of the December 1996 observation the 
line equivalent width decreases with continuum flux, although the
line flux and continuum flux are still correlated.

\begin{table*}
\centering
\caption{\label{diskline}Disk line model fits to flux 
selected  spectra}
\begin{tabular}{lccccccc}
  mean flux          & photon index&$EW_{\rm narrow}$&$f_{\rm disk}$ & $ EW_{\rm disk}$& $i$ & $-q$ & $\chi^2$ (dof) \\
$[\rm{erg\; s^{-1}\; cm^{-2}}]$ &  &$[{\rm eV}]$     &$[10^{-5} {\rm ph\; cm^{-2}s^{-1}}] $&$[{\rm eV}]$    & [deg] &  & \\[10pt]
\multicolumn{6}{c}{Long term momitoring} \\[10pt]
 $5.46\cdot10^{-12}$ &$1.60\pm0.10$&397&$1.93\pm0.89 $&$192\pm80$ & $0\pm90$ & $2-3.3$ & 18.5(44) \\
 $1.62\cdot10^{-11}$ &$1.89\pm0.04$&162&$7.19\pm1.49 $&$389\pm81$ & $27\pm7$ & $>3.8$ & 19.1(44)\\
 $2.77\cdot10^{-11}$ &$2.10\pm0.02$&105&$11.33\pm1.66 $&$367\pm54$ & $31\pm4$ & $>5.7$ & 29.7(44)\\
 $3.35\cdot10^{-11}$ &$2.16\pm0.03$&89 &$11.40\pm3.25$&$327\pm93$ & $28\pm7$ & $>2.9$ & 18.0(44)\\
 $4.06\cdot10^{-11}$ &$2.19\pm0.03$&80 &$9.74\pm1.90$&$216\pm42$ & $24\pm7$ & $>5.2$ & 22.4(44)\\
 $5.21\cdot10^{-11}$ &$2.29\pm0.03$&61 &$10.56\pm3.17$&$204\pm61$ & $32\pm9$ & $>4.8$ & 26.1(44)\\
 $6.74\cdot10^{-11}$ &$2.33\pm0.03$&49 &$16.84\pm3.24$&$181\pm35$ & $0.\pm19$& $>2.8$ &23.6(44)\\[10pt]
\multicolumn{6}{c}{December 1996} \\[10pt]
 $6.78\cdot10^{-12}$ &$1.81\pm0.09$&391&$5.31\pm1.12 $&$675\pm142$ & $28\pm6$& $>7.0 $&45.0(44) \\
 $1.53\cdot10^{-11}$ &$1.86\pm0.06$&186&$10.07\pm2.17 $&$566\pm121$ & $25\pm5$& $>5.2$ &33.3(44) \\
 $2.64\cdot10^{-11}$ &$2.11\pm0.04$&120&$14.03\pm2.07 $&$458\pm68$ & $23\pm5$ & $>11.2$&28.3(44) \\
 $3.61\cdot10^{-11}$ &$2.23\pm0.06$& 91&$10.64\pm4.66 $&$267\pm117$ & $24\pm15$&$>3.2$ &41.8(44) \\
\end{tabular}
\end{table*}

\section{DISCUSSION}

We show that the variable X-ray spectrum of NGC~4051 can be well
described by a model that includes two principal components:

\begin{itemize}
\item[1.] A hard, constant, component including a narrow iron 
fluorescence line as revealed during the extreme low state in May
1998.

\item[2.] A variable component including a power law with strongly
variable slope and a very broad emission line. We find a strong
correlation between the power law slope and the source flux.  On the
long time-scales of the monitoring observations the line flux is
correlated with the continuum flux, although not in a simple manner.
The reflected fraction of the disk component is less than $R=1$ at all
flux levels.

\end{itemize}

This two-component model is consistent with the model we proposed for
the X-ray spectral variability of the Seyfert 2 galaxy NGC~5506
\cite{Lamer2000}. 

We note that even if the constant hard component is not included in
our spectral fits, the general form of our results is not
significantly affected, as the assumed constant component is
relatively weak. Indeed, the lack of inclusion of a hard continuum
component would result in an even larger spectral index variation with
flux.

\subsection{The Iron Line}

Even on the long timescales probed by our time-averaged monitoring data,
the flux-dependent behaviour of the broad iron line is complex.  For
these data, the line flux increases more-or-less
proportionally with the continuum flux, resulting in a roughly
constant equivalent width over a decade range of flux.  The December
1996 long-look data also show line flux increasing with continuum
flux, although the relation is not directly proportionate, so that the
equivalent width decreases with flux.  In fact, the line fluxes
measured in December 1996 are consistently larger than the
corresponding fluxes measured from the long-term monitoring data.
The anti-correlation of line equivalent width and continuum flux in 
December 1996 timescales is also in  contrast  to the result of \cite {Wang}, 
who report a positive  correlation during an {\it ASCA} observation in 1994.

The discrepancy between the iron line behaviour in the December 1996
and long-term monitoring data might be explained if there is
additional short-term variability in the iron line which is not simply
related to the continuum flux. 
For example, Vaughan \& Edelson (2001)
show that in the Seyfert~1 MCG-6-30-15, the broad iron line flux
varies significantly but independently of short term continuum
variations.

  One possibility is that the iron line flux tracks the
long-term variations in the continuum flux (which are being probed to
some extent with the long-term monitoring data), but responds only
weakly to the short-term variations which are observed during the
December 1996 long-look observation. A number of theoretical papers
have been written to explain why the iron line flux may not
vary linearly with the continuum flux (eg \ncite{Matt}, \ncite{Nayakshin},
\ncite{Ballantyne}),
often involving ionised discs but these models have so far been
largely untroubled by data. We are aquiring more long-term
monitoring data, sampling a broader range of long-term flux
variations,  to determine whether the iron line does follow the
continuum on long timescales and to provide some constraints for
theoretical models.

%Long look observations with 
%{\em XMM-Newton} will be suitable to make more accurate measurements
%of the  iron line variability 
%at shorter time-scales and will certainly clarify the picture.

\subsection{The Reflected Component}

Our observations of NGC~4051 do not support the correlation between
the photon index and the reflected fraction $R$ in the {\sc pexrav}
model as reported from {\em ASCA} spectroscopy of a sample of Seyfert
galaxies \cite{Zdziarski} .  From Fig. \ref{dec_cont} it is obvious
that the reflected fraction remains below $R=1$ even for the softest
states of the source. There is also no evidence for this correlation
in the {\em RXTE} spectra of NGC~5506 \cite{Lamer2000}. We therefore
suggest that the reported correlation does not apply to the variations
of photon index and reflected fraction in a given object.

\subsection{Continuum Spectral Variability}

During our monitoring campaign the primary continuum 2-10 keV photon
index, $\Gamma$, varies strongly with photon flux from 1.60 at the
lowest flux levels to 2.35 at the highest fluxes.  The same
correlation is also observed on the shorter time-scales of the
December 1996 long look. Flux-$\Gamma$ correlations have been observed
before in NGC~4051 \cite{Matsuoka} and other Seyfert galaxies
(eg NGC~4151, \ncite{Perola}) but, as in the present paper, it has
only recently been possible to disentangle the effects of reflection
and variations of the primary X-ray spectrum (eg \ncite{Lamer2000},
\ncite{Chiang}, \ncite{Lee2000}).

The slope-luminosity correlation is often explained by stronger
cooling of the accretion disk corona during episodes of high thermal
seed photon flux from the accretion disk itself (eg \ncite{Pietrini},
\ncite{Malzac}).  \scite{Haardt} have calculated luminosity -- spectral
index relations in Compton cooled accretion disk coronae.  For a
compact, pair dominated corona they predict spectral index variations
of $\Delta\Gamma\sim 0.3$ for luminosity variations by more than a
factor of 20. However the variations seen here exceed their
predictions and imply, in their scenario, a non-pair dominated corona.
In certain regimes this model predicts a positve correlation of 2-10 keV flux 
and spectral hardness, which is not observed in NGC~4051.
\scite{Pietrini} point out that the spectral index depends almost solely on
the ratio of seed photon compactness $l_s$ and hot plasma heating rate compactness
$l_h$ with $\alpha=1.6(l_s/l_h)^{1/4}$. The observed spectral indices in NGC~4051 
then correspond to $l_s/l_h=0.02..0.4$.

Examination of fig~\ref{lineflux} shows that the change of spectral
index with flux is not linear. The rate of increase of spectral index
with flux is very rapid at low fluxes but decreases at the highest
fluxes where the spectral index approaches an asymptotic level.  This
saturation of the `spectral index/flux' relationship has been known
for some time; eg the saturation was clearly visible in our early RXTE
monitoring observations of MCG-6-30-15 and was reported by
\scite{mch98} where it was suggested that the
relationship might derive from the combination of a constant spectrum
hard component, and a steeper spectrum variable component.

Saturation of the spectral index/flux relationship was again reported
in MCG-6-30-15 by \scite{Shih} from a long ASCA observation. In
MCG-6-30-15 both the long term RXTE monitoring and short term ASCA
observations agree that the saturation level of the spectral index is
$\sim2.1$ (see fig 7 of \ncite{mch98} and fig 8 of \ncite{Shih}). 
However the monitoring observations cover a wider flux and
spectral range and show variation of the spectral index between 1.65
and 2.05 whereas the continuous ASCA observation only shows an index
variation between 1.9 and 2.1. Similarly in NGC~4051
(fig~\ref{lineflux}) we see that the monitoring observations cover a
wider flux and spectral range than the December 1996 long look and, as
with MCG-6-30-15, the resultant time-averaged spectral index/flux
relationship is smoother. We note, however, that although the lowest
spectral index so far measured in the RXTE monitoring observations is
about the same in both NGC~4051 and MCG-6-30-15, the saturation level
is $\sim2.4$ in NGC~4051 compared to $\sim2.1$ in MCG-6-30-15.

In \scite{mch98} we suggested that the torus might be the source of the
possible hard, constant, component. However the very hard spectral
component found in the May 1998 very low state, which represents an
upper limit to the torus contribution, was removed before producing
the spectral index/flux relationship (fig~\ref{lineflux}). Thus, if we
wish to retain the two-component spectral model, we require a
different location for the bulk of the hard component. However any
hard component produced closer to the central source by reprocessing
of primary radiation is likely to vary, although probably not as
rapidly as the primary continuum. Our time-averaged spectra wash out
short term variability so a hard component varying slowly in line with
the primary continuum could not produce the observed spectral index
flux relationship.  Unless a hard component is produced by some
mechanism other than reprocessing it is therefore more likely that the
change of spectral index with flux is driven by some change in the
physical properties of the primary continuum source, eg by changes in the
seed photon populations as discussed above.

Any such physical model of the primary continuum must be able to
produce different saturation levels in different sources and must be
able to account for both the wide spectral range, and the very hard
spectra measured at the lowest flux levels, in our monitoring
observations.

\label{lastpage}

\end{document}